\documentclass[conference,letterpaper]{IEEEtran}

\usepackage{graphicx}
\usepackage{color}

\def\ve#1{{\mathchoice{\mbox{\boldmath$\displaystyle #1$}}%
              {\mbox{\boldmath$\textstyle #1$}}%
              {\mbox{\boldmath$\scriptstyle #1$}}%
              {\mbox{\boldmath$\scriptscriptstyle #1$}}}}

\newcommand{\evs}{\mathcal{E}}

\newcommand{\Ex}{E}

\newcommand{\SNRdB}{10\log_{10}(\bar{E}_b / \mathcal{N}_0)}
\newcommand{\diag}{\mathrm{diag}}
\newcommand{\PEP}{\mathrm{PEP}}
\newcommand{\No}{\mathcal{N}_0}

\newcommand{\sst}{-5mm}
\newcommand{\ssf}{-6mm}

\graphicspath{{./figures/}}

\begin{document}

\title{Error Rate Analysis for Coded Multicarrier Systems over Quasi-Static Fading Channels}

\author{\authorblockN{Chris Snow, Lutz Lampe, and Robert Schober}
        \authorblockA{Department of Electrical \& Computer Engineering\\
                        University of British Columbia, Vancouver, Canada\\
                        \{csnow,lampe,rschober\}@ece.ubc.ca}}

\maketitle

\begin{abstract}
This paper presents two methods for approximating the performance of coded multicarrier systems operating over frequency-selective, quasi-static fading channels with non-ideal interleaving. The first method is based on approximating the performance of the system over each realization of the channel, and is suitable for obtaining the \textsl{outage} performance of this type of system. The second method is based on knowledge of the correlation matrix of the frequency-domain channel gains and can be used to directly obtain the average performance. Both of the methods are applicable for convolutionally-coded interleaved systems employing Quadrature Amplitude Modulation (QAM). As examples, both methods are used to study the performance of the Multiband Orthogonal Frequency Division Multiplexing (OFDM) proposal for high data-rate Ultra-Wideband (UWB) communication.
\end{abstract}

\section{Introduction}
\label{sec:intro}

Multicarrier communications systems based on Orthogonal Frequency Division Multiplexing (OFDM) have gained interest from the communications community in recent years, as evidenced by wireless standards such as 802.11~a/g and high-rate Ultra-Wideband (UWB)~\cite{802.11,MBOFDM}. In many situations, the wireless channel for these systems can be assumed very slowly time-varying relative to the transmission rate of the device, and can be approximated as quasi-static for the duration of one or more packet transmissions. As well, since most OFDM devices use a relatively large bandwidth, the channel is frequency-selective. This motivates an interest in the analysis of the performance of coded multicarrier systems when transmitting over quasi-static frequency-selective fading channels. Throughout this paper, we will assume that the OFDM system is designed such that the cyclic prefix is always longer than the channel impulse response. Thus, we can consider the channel in the frequency domain, with one channel gain per subcarrier and a correlation matrix representing the correlation between the gains of different subcarriers.

When considering the performance of systems over quasi-static channels, the concept of \textsl{outage} is often used. That is, since (in a packet-based transmission system) each packet will be transmitted over only one realization of the quasi-static channel, we consider some channel realizations to be in outage (that is, they do not support the required data rate). The worst-case performance of the non-outage cases is then studied, providing information about what minimum performance can be expected of the system given a certain allowable outage rate.

It should be emphasized that classical bit error rate (BER) analysis techniques for coded systems~\cite{Proakis,CaireEtAl98} are not applicable in this setting because the channel is (a) non-ideally interleaved (resulting in non-zero correlation between adjacent coded bits), and (b) quasi-static (which limits the number of distinct channel gains to the number of OFDM tones). 

\textsl{Related Work:} In \cite{ML99}, Malkam\"aki and Leib consider the performance of convolutional codes with non-ideal interleaving over block fading channels. They make use of the generalized transfer function (GTF) \cite{LWK93} %,RM95} 
in order to obtain the pairwise error probability in the case when the channel gains for each transmitted symbol are not unique (due to a small number of fading gains as a result of the block fading model). This approach is similar in some ways to that presented in Section~\ref{sec:rba}. The major difference is that our method does not require the GTF of the code, which may be difficult to obtain~\cite{CCT04}. %. (Evaluation of the GTF for given codes is considered in e.g. \cite{CCT04}.) 
Instead, we apply the concept of error vectors, introduced in Section~\ref{sec:evs}.

In \cite{V01}, Veeravalli presents a general approach for analyzing the performance of uncoded diversity-combining systems on correlated fading channels. In particular, the probability of error is expressed (for Rayleigh-fading channels) as a function of the complex covariance matrix of the channel gains. In Section~\ref{sec:avg} we extend this approach to handle coded multicarrier systems employing arbitrary Quadrature Amplitude Modulation (QAM) modulation schemes.

\textsl{Organization:} Section~\ref{sec:sysmodel} introduces the system model considered in this paper. In Section~\ref{sec:analysis}, we present two approaches to performance analysis: a per-realization based approach for system performance evaluation which readily leads to estimation of outage-based performance measures, and an approach, based on the channel correlation matrix, which can be used to obtain the average system performance. Numerical results illustrating the effectiveness of both proposed methods are presented in Section~\ref{sec:results}. Finally, Section~\ref{sec:conclusions} concludes the paper.

\section{System Model}
\label{sec:sysmodel}

Throughout this paper we consider an $N$-tone OFDM system, using $M$-ary QAM (M-QAM) on each subcarrier. The system employs a punctured convolutional code of rate $R_c$. 

We assume that the transmitter selects a vector of $R_cN\log_2M$ random message bits for transmission, denoted by $\ve{b} = [b_1\;b_2 \dots b_{R_cN\log_2M}]^T$ (where $[\cdot]^T$ denotes vector transposition). The vectors $\ve{c}$ and $\ve{c}^\pi$ of length $L_c=N\log_2M$ represent the bits after encoding/puncturing and after interleaving, respectively. The bits $\ve{c}^\pi$ are then modulated using $M$-QAM on each subcarrier. The resulting $N$ modulated symbols are denoted by the vector $\ve{x} = [x_1\;x_2\;\dots\;x_N]^T$.

The symbols $\ve{x}$ are transmitted through a quasi-static fading channel with frequency-domain channel gains $\ve{h} = [h_1\;h_2\;\dots\;h_N]$. Writing $\ve{H}=\diag(\ve{h})$, where $\diag(\ve{h})$ denotes a matrix with the elements of $\ve{h}$ on the main diagonal, we can express the received symbols $\ve{r}$ as
\begin{equation}
\ve{r} = \sqrt{E_s}\ve{H}\ve{x} + \ve{n}\;,
\end{equation}
where $\ve{n}$ is a vector of independent complex additive white Gaussian noise (AWGN) variables with variance $\mathcal{N}_0$ and $E_s$ is the energy per modulated symbol. The energy per information bit is $E_b = E_s / (R_c\log_2M)$.

We assume perfect timing and frequency synchronization. The receiver employs a soft-output detector followed by deinterleaving, depuncturing, and Viterbi decoding, resulting in an estimate $\hat{\ve{b}} = [\hat{b}_1\;\hat{b}_2\;\dots\;\hat{b}_{R_cN\log_2M}]^T$ of the original transmitted information bits.
%After demodulation, the received coded bits are given by $\ve{v}^\pi$, and after deinterleaving by $\ve{v}$. We also term $\ve{z}$ as the vector of modulated symbols corresponding to $\ve{v}^\pi$. After depuncturing and decoding via the Viterbi algorithm, we obtain an estimate $\hat{\ve{b}} = [\hat{b}_1, \hat{b}_2, \dots, \hat{b}_{R_cN\log_2M}]^T$ of the original transmitted information bits. 

\section{Two Performance Analysis Methods}
\label{sec:analysis}

In this section, we present two methods for approximating the performance of coded multicarrier systems operating over frequency-selective, quasi-static fading channels. The first method is based on approximating the performance of the system over individual channel realizations, and is suitable for obtaining the outage performance. The second method is based on knowledge of the correlation matrix of the frequency-domain channel gains and can be used to directly obtain the average performance. Both methods are based on considering the set of error vectors, introduced below.

One major problem in the analysis of $M$-QAM modulation schemes with $M>4$ is that the probability of error for a given bit depends on the whole transmitted symbol (i.e. it also depends on the other bits in the symbol). For this reason, for the combination of convolutional coding and $M$-QAM it is not sufficient to adopt the classical approach of considering deviations from the all-zero codeword. In theory, one must average over all possible choices for $\ve{c}$. Since this is computationally intractable, we simply assume the transmitted information bits $\ve{b}$ (and hence $\ve{x}$) are chosen randomly. For 2-QAM and 4-QAM (where the joint linearity of code and modulator is maintained) this is exactly equivalent to considering an all-zero codeword. In the case of $M>4$, we have verified for the two analysis methods proposed below that the results are practically invariant to the choice of $\ve{b}$.

\subsection{Set of Error Vectors}
\label{sec:evs}

Let $\evs$ be the set of all $L$ vectors  $\ve{e}_j$ $(1{\le}j{\le}L)$ of code output (after puncturing) associated with input sequences with Hamming weight less than $w_{\mathrm{max}}$, i.e., $\evs = \left\{\ve{e}_1,\ve{e}_2,\dots,\ve{e}_L\right\}$. Let $l_j$ and $a_j$ be the length of $\ve{e}_j$ and the number of information bit errors associated with $\ve{e}_j$, respectively. As with standard union-bound techniques for convolutional codes~\cite{Proakis}, the low-weight terms will dominate the error probability. Hence it is sufficient to choose a small $w_{\mathrm{max}}$ --- for example, the punctured code of rate $1/2$ considered in Section~\ref{sec:results} has a free distance of $9$, and choosing $w_{\mathrm{max}}=14$ (resulting in a set of $L=242$ error vectors of maximum length $l=60$) provides results virtually identical to those obtained using larger $w_{\mathrm{max}}$ values.

In the following, we refer to $\ve{e}_j$ as an ``error vector'' and to $\evs$ as the set of error vectors. Note that $\evs$ can be straightforwardly obtained from the transfer function of the code, without resorting to the GTF approach as in~\cite{ML99}. It is also independent of the number of distinct channel gains (or the number of blocks in the context of~\cite{ML99}).

\subsection{Pairwise Error Probability}

We consider error events starting in a given position $i$ of the codeword ($1 \le i \le L_c$). We consider each error vector $\ve{e}_j$ for $1 \le j \le L$, and form the full error codeword 
\begin{equation}
\label{eq:q}
\ve{q}_{i,j} = [\underbrace{0\phantom{_j}\;0\;\dots\;0}_{i-1}\;\underbrace{\ve{e}_j}_{l_j}\;\underbrace{0\phantom{_j}\;0\;\dots\;0}_{L_c-l_j-i+1}]^T
\end{equation}
of length $L_c$ by padding $\ve{e}_j$ with zeros on both sides as indicated above. Given the error codeword $\ve{q}_{i,j}$ and given that codeword $\ve{c}$ is transmitted, the competing codeword is given by
%  By the linearity of the code, the vector $\ve{v}_{i,j}$ bits given the $j^{\mathrm{th}}$ error vector starting in the $i^{\mathrm{th}}$ position.
%
\begin{equation}
\label{eq:v}
\ve{v}_{i,j} = \ve{c} \oplus \ve{q}_{i,j}
\end{equation}
where $\oplus$ denotes XOR. Letting $\ve{z}_{i,j}$ be the vector of QAM symbols associated with $\ve{v}_{i,j}^\pi$ (the interleaved version of $\ve{v}_{i,j}$), the pairwise error probability (PEP) for the $j$th error vector starting in the $i$th position is
%\noindent of length $L_c$ by padding $\ve{e}_j$ with zeros on both sides as indicated above. By the linearity of the code
%
%\begin{equation}
%\label{eq:v}
%\ve{v}_{i,j} = \ve{c} \oplus \ve{q}_{i,j}
%\end{equation}
%
%where $\oplus$ denotes XOR, i.e., $\ve{v}_{i,j}$ is the vector of received bits given the $j^{\mathrm{th}}$ error vector starting in the $i^{\mathrm{th}}$ position.
%Letting $\ve{z}_{i,j}$ be the vector of received symbols associated with the received coded bits $\ve{v}_{i,j}^\pi$, the pairwise error probability (PEP) for the $j$th error vector starting in the $i$th position is given by
%
\begin{equation}
\label{eq:PEP}
\PEP_{i,j} = Q\left(\sqrt{\frac{E_s}{2\No}||\ve{H}(\ve{x}-\ve{z}_{i,j})||^2}\right) \;,
\end{equation}
where $Q(\cdot)$ is the Gaussian Q function~\cite{Proakis}.

\subsection{Per-Realization Performance Analysis (``Method I'')}
\label{sec:rba}

In this section, we obtain an approximation of the BER for a particular channel realization $\ve{H}=\diag(\ve{h})$, which we denote as $P(\ve{H})$. This method leads naturally to the analysis of the \textsl{outage} performance, which is a useful measure for schemes operating over quasi-static channels. For simplicity, we shall refer to this as Method I in the remainder of this paper.

The PEP for an error vector $\ve{e}_j$ ($1 \le j \le L$) with the error event starting in a position $i$ ($1 \le i \le L_c$) is given by (\ref{eq:PEP}). The corresponding bit error rate for this event is given by 
\begin{equation}
\label{eq:BERij}
P_{i,j}(\ve{H}) = a_j \cdot \PEP_{i,j}(\ve{H}) \;.
\end{equation}
Summing over all $L$ error vectors, we obtain an approximation of the BER for the $i^{\mathrm{th}}$ starting position as
\begin{equation}
\label{eq:BERi}
P_i(\ve{H}) = \sum_{j=1}^{L}a_j \cdot \PEP_{i,j}(\ve{H}) \;.
\end{equation}
We note that (\ref{eq:BERi}) can be seen as a standard truncated union bound for convolutional codes (i.e., it is a sum over all error events of Hamming weight less than $\omega_{\mathrm{max}}$). We also note that we can tighten this bound by limiting $P_i(\ve{H})$ to a maximum value of $1/2$ before averaging over starting positions~\cite{ML99}. Finally, since all starting positions are equally likely, the BER $P(\ve{H})$ can be written as 
\begin{equation}
\label{eq:BERgivenH}
P(\ve{H}) = \frac{1}{L_c}\sum_{i=1}^{L_c}\min \left[ \frac{1}{2}, \sum_{j=1}^{L} P_{i,j}(\ve{H}) \right]\;.
\end{equation}
Table~\ref{table:pseudo_rba} contains pseudocode to calculate $P(\ve{H})$ according to (\ref{eq:BERgivenH}). This method readily leads to the consideration of the outage performance. We evaluate (\ref{eq:BERgivenH}) for many channel realizations. For a given $X$\% outage rate, the worst-performing (100-$X$)\% of realizations are considered in outage (i.e., they are not capable of supporting the required data rate), and the worst-case performance of the non-outage cases is shown. This provides information about the minimum performance that can be expected of the system given the $X$\% outage rate.

We should note here that (\ref{eq:BERgivenH}) does not take into account edge effects at the end of a coded OFDM symbol (i.e., near the end of a symbol not all error vectors represent valid errors because the codeword may end before the error vector does). It is possible to modify (\ref{eq:BERgivenH}) in order to disregard error events that do not ``fit'' within the symbol length. However, such modification makes a minimal difference in the final BER of the overall system.

\begin{table}[t]
\caption{\label{table:pseudo_rba}Pseudocode for Method I. Final BER is $P$ (for given $\ve{H}$).}
\vspace{-3mm}
\centerline{\begin{tabular}{ll}
\hline
\vspace{-2mm} & \\
\tt 1  & \tt $P$ := 0\\
\tt 2  & \tt for $i$ := 1 to $L_c$ do \\
\tt 3  & \tt ~~$P_i$ := 0 \\
\tt 4  & \tt ~~for $j$ := 1 to $L$ \\
\tt 5  & \tt ~~~~form $\ve{q}_{i,j}$ as per (\ref{eq:q})\\
\tt 6  & \tt ~~~~form $\ve{v}_{i,j}$ as per (\ref{eq:v})\\
\tt 7  & \tt ~~~~form $\ve{v}^\pi_{i,j}$ and $\ve{z}_{i,j}$ from $\ve{v}_{i,j}$\\
\tt 8  & \tt ~~~~calculate $P_{i,j}$ as per (\ref{eq:BERij})\\
\tt 9  & \tt ~~~~$P_i$ := $P_i$ + $P_{i,j}$\\
%\tt 10  & \tt ~~endfor \\
\tt 10 & \tt ~~$P$ := $P$ + min($\frac{1}{2}$,$P_i$)\\
%\tt 12 & \tt endfor \\
\tt 11 & \tt $P$ := $P$ / $L_c$ \\
\vspace{-2mm} & \\
\hline
\end{tabular}}
\vspace{\sst}
%\vspace{3.90mm}
\end{table}

\subsection{Averaging Approach (``Method II'')}
\label{sec:avg}

In this section, we propose a method, based on knowledge of the frequency-domain channel correlation matrix, which can be used directly in order to obtain the average BER performance of coded multicarrier systems. For simplicity, we refer to this as Method II in the remainder of this paper.

For this method we will explicitly assume that the elements of $\ve{h}$ are Rayleigh-distributed and have known correlation matrix $\ve{\Sigma_{hh}}$ (in practice, $\ve{\Sigma_{hh}}$ can be obtained from actual channel measurements, or can be numerically estimated by measuring the correlation in many channel realizations from a given model).

Note that for an error vector $\ve{e}_j$ starting in position $i$, only the non-zero terms of $(\ve{x}-\ve{z}_{i,j})$ contribute to the PEP. Let $\ve{x'}$, $\ve{z'}_{i,j}$ and $\ve{H'}=\diag(\ve{h'})$ represent the transmitted symbols, received symbols, and channel gains corresponding to the non-zero distances of  $(\ve{x}-\ve{z}_{i,j})$, respectively, and form $\ve{R_{h'h'}}$ by extracting the elements from $\ve{\Sigma_{hh}}$ which correspond to $\ve{h'}$. Let $\ve{D} = \diag(\ve{x'}-\ve{z'}_{i,j})$ be the diagonal matrix of non-zero distances. Now $\ve{g} = \ve{H'}(\ve{x'}-\ve{z'}_{i,j}) =  \ve{D}\ve{h'}$. We have 
\begin{eqnarray}
\Ex(\ve{g}) &=& \ve{0} \;,\nonumber \\
\Ex(\ve{g}\ve{g}^H) &=& \ve{R_{gg}} = \ve{D}\ve{R_{h'h'}}\ve{D}^H \;,\nonumber
\end{eqnarray}
where $\Ex(\cdot)$ denotes expectation and $(\cdot)^H$ denotes Hermitian transpose. The distribution of $\ve{g}$ is zero-mean complex Gaussian with covariance matrix $\ve{R_{gg}}$. Following \cite[Eq. (7)]{V01}, we can write the bit error probability for the $j^\mathrm{th}$ error vector starting in the $i^\mathrm{th}$ position as 
\begin{equation}
\label{eq:ProbInt}
\bar{P}_{i,j} = \frac{a_j}{\pi}\int_{0}^{\pi/2}\left[ \det\left( \frac{E_s\ve{R_{gg}}}{\No\sin^2\theta} + \ve{I}\right) \right]^{-1}d\theta\;.
\end{equation}
\noindent Note that (\ref{eq:ProbInt}) can be written in closed form in terms of the eigenvalues of $\ve{R_{gg}}$, as in \cite{V01}. Alternatively, (\ref{eq:ProbInt}) can simply be solved directly via standard numerical integration techniques. The latter approach is favorable in the case we are considering, since $\ve{R_{h'h'}}$ (and thus $\ve{R_{gg}}$) is in general different for each $i,j$, necessitating a new eigendecomposition.

Now, summing over all $L$ error vectors, the BER for the $i^{\mathrm{th}}$ starting position can be written as
\begin{equation}
\bar{P}_i = \sum_{j=1}^{L}\bar{P}_{i,j} \;.
\end{equation}
Finally, since all starting positions are equally likely to be used, the average BER $\bar{P}$ can be written as 
\begin{equation}
\label{eq:BERavg}
\bar{P} = \frac{1}{L_c}\sum_{i=1}^{L_c}\bar{P}_i = \frac{1}{L_c}\sum_{i=1}^{L_c} \sum_{j=1}^{L} \bar{P}_{i,j} \;.
\end{equation}
Table~\ref{table:pseudo_corr} contains pseudocode to calculate $\bar{P}$ according to (\ref{eq:BERavg}). 

\begin{table}[t]
\caption{\label{table:pseudo_corr}Pseudocode for Method II. Final BER is $\bar{P}$.}
\vspace{-2.5mm}
\centerline{\begin{tabular}{ll}
\hline
\vspace{-2mm} & \\
\tt 1  & \tt $\bar{P}$ := 0\\
\tt 2  & \tt for $i$ := 1 to $L_c$ do \\
\tt 3  & \tt ~~for $j$ := 1 to $L$ \\
\tt 4  & \tt ~~~~form $\ve{q}_{i,j}$ as per (\ref{eq:q}) \\
\tt 5  & \tt ~~~~form $\ve{v}_{i,j}$ as per (\ref{eq:v})\\
\tt 6  & \tt ~~~~form $\ve{v}^\pi_{i,j}$ and $\ve{z}_{i,j}$ from $\ve{v}_{i,j}$\\
\tt 7  & \tt ~~~~form $\ve{x'}_{i,j}$, $\ve{z'}_{i,j}$, $\ve{h'}_{i,j}$ and $\ve{R_{h'h'}}$ \\
\tt 8  & \tt ~~~~Compute $\ve{D} := \diag(\ve{x'}-\ve{z'}_{i,j})$ \\
\tt 9  & \tt ~~~~Compute $\ve{R_{gg}} := \ve{D}\ve{R_{h'h'}}\ve{D}^H$ \\
\tt 10 & \tt ~~~~Compute (\ref{eq:ProbInt}) to obtain $\bar{P}_{i,j}$ \\
\tt 11 & \tt ~~~~$\bar{P}$ := $\bar{P}$ + $\bar{P}_{i,j}$\\
%\tt 12 & \tt ~~endfor \\
%\tt 13 & \tt endfor \\
\tt 12 & \tt $\bar{P}$ := $\bar{P}$ / $L_c$ \\
\vspace{-2mm} & \\
\hline
\end{tabular}}
\vspace{\sst}
%\vspace{1mm}
\end{table}

\section{Numerical Results}
\label{sec:results}

In this section, we present numerical results illustrating the performance analysis methods presented in Section~\ref{sec:analysis}. 

\subsection{System and Channel Model}

As an example OFDM system, we have chosen the Multiband OFDM system (MB-OFDM) proposed for IEEE 802.15 TG3a High Data Rate UWB~\cite{MBOFDM}. The MB-OFDM system uses 128 tones and operates by hopping over 3 sub-bands (one hop per OFDM symbol) in a predetermined pattern. We will assume that hopping pattern 1 of \cite{MBOFDM} is used (i.e. the sub-bands are hopped in order). As a result we can consider MB-OFDM as an equivalent 384 tone OFDM system. After disregarding pilot, guard, and other reserved subcarriers, we have $N=300$ data-carrying tones. 

Channel coding in the proposed standard consists of 
%classical bit-interleaved coded modulation (BICM) \cite{CaireEtAl98} with 
 a punctured maximum free distance rate $1/3$ constraint length 7 convolutional encoder and a multi-stage block-based interleaver (see \cite{MBOFDM} for details). After modulation, modulated symbols are optionally repeated in time (in two consecutive OFDM symbols) and/or frequency (two tones within the same OFDM symbol), reducing the effective code rate by a factor of 2 or 4 and providing an additional spreading gain for low data rate modes. In the framework of our analysis, we can equivalently consider this time/frequency spreading as a lower-rate convolutional code with repeated generator polynomials. In the proposed standard, the interleaved coded bits are mapped to quaternary phase-shift keying (QPSK) symbols using Gray labeling. As an extension for higher data rates, we also consider Gray-labeled 16-QAM.

For a meaningful performance analysis of the MB-OFDM proposal, we consider the channel model developed under IEEE 802.15 for UWB systems \cite{MolischEtAl03}.
The channel impulse response is a based on a modified Saleh-Valenzuela model~\cite{Saleh+Valenzuela87}. 
%Multipath rays arrive in clusters, with exponentially distributed cluster and ray interarrival times.Both clusters and rays have decay factors chosen to meet a given power decay profile. The ray amplitudes are modeled as lognormal random variables, and each cluster of rays also undergoes a lognormal fading. To provide a fair system comparison, the total multipath energy is normalized to unity. 
As well, the entire impulse response undergoes an ``outer'' lognormal shadowing.  The channel impulse response is assumed time invariant during the transmission period of (at least) one packet (see \cite{MolischEtAl03} for a detailed description). We consider the UWB channel parameter set referred to as CM1~\cite{MolischEtAl03}.

The OFDM transmit signal experiences a frequency non-selective fading channel with fading along the frequency axis. Thereby, the outer lognormal shadowing term mentioned above is irrelevant for the fading characteristics as it  affects all tones equally. Denoting the lognormal term by $G$, we obtain the corresponding \textsl{normalized} frequency-domain fading coefficients as 
\begin{equation}
\label{eq:normchannel}
        \ve{h}^n = \ve{h}/G\;.
\end{equation}
The elements of $\ve{h}^n$ are well-approximated as zero-mean complex Gaussian random variables~\cite{SLS05a}. This allows us to apply the analysis of Section~\ref{sec:avg} to the UWB channel without lognormal shadowing, and then average over the lognormal shadowing distribution in order to obtain the final system performance over the composite UWB channel. We note that this is only relevant for Method II --- for Method I the distribution of $\ve{h}$ is not important.

\subsection{Realization-based Approach}

\begin{figure}[t]
\centerline{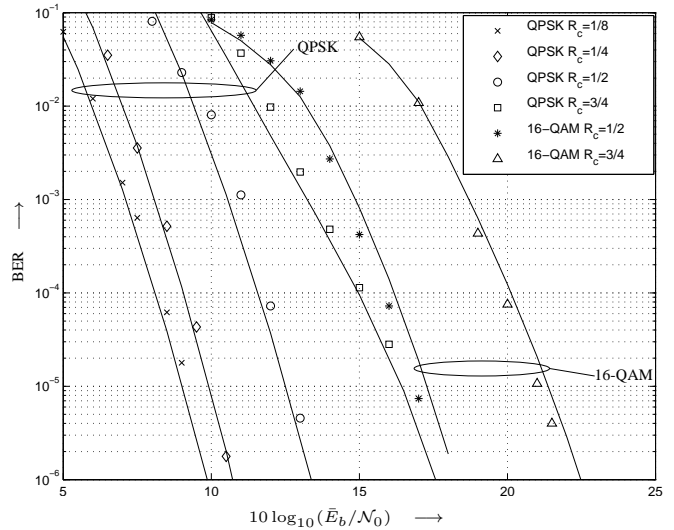}
\vspace{-1mm}
\caption{\label{fig:realiz}10\% Outage BER vs. $10\log_{10}(\bar{E}_b / \mathcal{N}_0)$ from Method I (lines) and simulation results (markers) for different code rates and modulation schemes. UWB CM1 channel with lognormal shadowing. Code rates $1/4$ and $1/8$ include repetition.}
\vspace{\ssf}
%\vspace{4mm}
\end{figure}

In Figure~\ref{fig:realiz} we present the 10\% outage BER as a function of $\SNRdB$ (the signal-to-noise ratio per information bit) obtained using Method I (lines), as well as simulation results (markers) for different code rates and modulation schemes using a set of 100 UWB CM1 channel realizations with lognormal shadowing. The 10\% outage BER is a common performance measure in UWB systems, cf. e.g.~\cite{MBOFDM}. We can see that Method I is able to accurately predict the BER for QPSK and 16-QAM modulation schemes with a variety of different code rates, with a maximum error of less than 0.5~dB. It is also important to note that obtaining the Method I result requires significantly less computation than is required to obtain the simulation results for all 100 UWB channel realizations. We should also note that it is not clear that even 100 channel realizations (standard for UWB system performance analysis) is sufficient to accurately capture the true system performance (see Section~\ref{sec:warning} for further discussion of this issue).

\subsection{Correlation-based Approach}

\begin{figure}[t]
\centerline{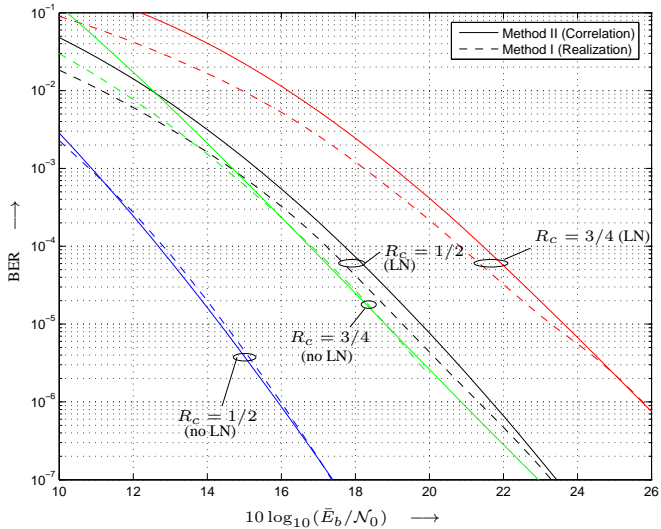}
\vspace{-1mm}
\caption{\label{fig:corr}Average BER vs. $10\log_{10}(\bar{E}_b / \mathcal{N}_0)$. Solid lines: correlation-based approach (Method II). Dashed lines: realization-based approach (Method I) for comparison. UWB CM1 channel, with and without lognormal shadowing (``LN'' and ``no LN'', respectively). 16-QAM, code rates $1/2$ and $3/4$.}
\vspace{\ssf}
%\vspace{2mm}
\end{figure}

In Figure~\ref{fig:corr} we present the average BER as a function of $\SNRdB$ for different codes rates with 16-QAM modulation over UWB channel CM1. The solid lines are the average BER obtained directly from the method of Section~\ref{sec:avg}, while the dashed lines are comparisons obtained from applying the method of Section~\ref{sec:rba} to 1000 channel realizations and averaging the results. We show curves both with and without the ``outer'' lognormal shadowing of the UWB channel model. As can be seen from Figure~\ref{fig:corr}, when the lognormal shadowing is neglected, the results obtained from the two methods show a close agreement. The difference between the two curves with lognormal shadowing is due to a slackening of the average bound which occurs when the results of Method II are averaged numerically over the ``outer'' lognormal distribution.

\subsection{A Note of Caution for System Designers}
\label{sec:warning}

\begin{figure}[t]
\centerline{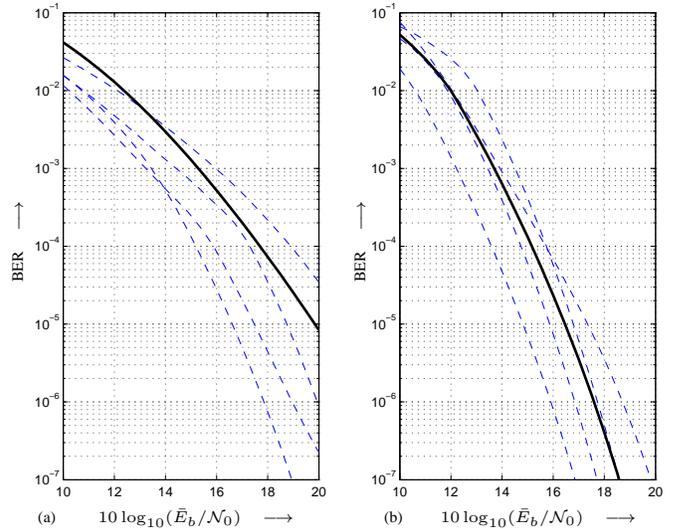}
\vspace{-1mm}
\caption{\label{fig:diff_avgandout}(a) Average BER vs. $10\log_{10}(\bar{E}_b / \mathcal{N}_0)$ for four different sets of 100 channels using Method I (dashed lines). For comparison: average performance from Method II (solid line). (b) 10\% Outage BER vs. $10\log_{10}(\bar{E}_b / \mathcal{N}_0)$ for four different sets of 100 channels using Method I (dashed lines). For comparison: 10\% Outage BER for a set of 1000 channels (solid line). Code rate $1/2$, 16-QAM, UWB CM1 channel with lognormal shadowing.}
\vspace{\ssf}
\end{figure}

Due to the high complexity of simulating system performance over quasi-static channels (which necessitates a large number of simulations over many channel realizations), system designers are tempted to use a relatively small set of (for example) 100 channel realizations to estimate system performance. Figure~\ref{fig:diff_avgandout}(a) (dashed lines) indicates the average BER with respect to $\SNRdB$ for 4 different sets of 100 UWB CM1 channel realizations, obtained via Method I. For comparison, the average performance obtained via Method II is also shown (solid line). We can see that the average system performance obtained using sets of only 100 channel realizations depends greatly on the specific realizations which are chosen. Similarly, Figure~\ref{fig:diff_avgandout}(b) illustrates the 10\% outage BER with respect to $\SNRdB$ for 4 different sets of 100 UWB CM1 channel realizations, obtained via Method I. For comparison the 10\% outage BER obtained using a set of 1000 realizations is also shown. We see that the outage BER curves, while less variable than the average BER curves in Figure~\ref{fig:diff_avgandout}(a), are still quite dependent on the selected channel realization set.

Based on the results above, it seems that performance evaluation for systems operating on quasi-static channels using only small numbers of channel realizations may be prone to inaccurate results. This is one of the main strengths of Method I presented in Section~\ref{sec:rba}: the performance can easily be evaluated over any (large) number of channel realizations without resorting to lengthly simulations.

\section{Conclusions}
\label{sec:conclusions}

In this paper, we have presented two methods for evaluating the performance of convolutionally-coded multicarrier systems employing QAM and operating over frequency-selective, quasi-static, non-ideally interleaved fading channels. 
The realization-based method (``Method I'') presented in Section~\ref{sec:rba} estimates the system performance over each realization of a channel with an arbitrary fading distribution, and is suitable for evaluating the outage performance of systems.  The method presented in Section~\ref{sec:avg} (``Method II''), based on knowledge of the correlation matrix of the frequency-domain channel gains, allows for direct calculation of the average system performance over the ensemble of quasi-static Rayleigh fading channel realizations. Both methods provide accurate estimates of the system performance as demonstrated by the numerical results of Section~\ref{sec:results}, and are both much simpler to evaluate than attempting system simulations over large sets of channel realizations.

%\vspace{-1mm}
\section*{Acknowledgment}

This work has been supported in part by the National Science and Engineering Research Council of Canada (Grant CRDPJ 320552) and Bell University Laboratories, and in part by a Canada Graduate Scholarship.

\bibliography{IEEEabrv,local}
\bibliographystyle{IEEEtran}

\end{document}